\documentclass{Interspeech2024}
\usepackage{multirow}
\usepackage{caption}
\usepackage{subcaption}
\usepackage{pifont}

\interspeechcameraready

\title{One-Class Learning with Adaptive Centroid 
Shift for Audio Deepfake Detection}

\name{Hyun Myung}{Kim}
\name{Kangwook}{Jang}
\name{Hoirin}{Kim}

\address{School of Electrical Engineering, KAIST, South Korea}
\email{\{hmy.kim, dnrrkdwkd12, hoirkim\}@kaist.ac.kr}


\begin{document}

\maketitle
\begin{abstract}
As speech synthesis systems continue to make remarkable advances in recent years, the importance of robust deepfake detection systems that perform well in unseen systems has grown.
In this paper, we propose a novel adaptive centroid shift (ACS) method that updates the centroid representation by continually shifting as the weighted average of bonafide representations.
Our approach uses only bonafide samples to define their centroid, which can yield a specialized centroid for one-class learning. 
Integrating our ACS with one-class learning gathers bonafide representations into a single cluster, forming well-separated embeddings robust to unseen spoofing attacks.
Our proposed method achieves an equal error rate (EER) of 2.19\% on the ASVspoof 2021 deepfake dataset, outperforming all existing systems.
Furthermore, the t-SNE visualization illustrates that our method effectively maps the bonafide embeddings into a single cluster and successfully disentangles the bonafide and spoof classes.
\end{abstract}
\section{Introduction}
Speech synthesis systems such as text-to-speech (TTS) \cite{ren2020fastspeech} and voice conversion (VC) \cite{qian2019autovc} are evolving rapidly with the development of deep learning.
These systems are easily accessible to the public at a low cost and produce sophisticated synthetic speech that is indistinguishable from genuine human speech.
Despite its positive aspects, there is also the potential for misuse, such as using a deepfake voice for criminal purposes \cite{stupp2019fraudsters}.
To address and prevent such issues, research on audio deepfake detection (ADD) that distinguishes between bonafide and fake speech is essential.

The primary challenge in ADD is to enhance generalization ability ensuring effective detection of unseen synthesis systems.
Some studies \cite{liu2023asvspoof, muller2022does} have indicated that spoofing detection systems suffer significant performance degradation when facing unseen spoofing attacks.
To enhance generalization ability, recent studies have focused on data augmentation techniques such as reverberation \cite{cai2017countermeasures} and transmission effects \cite{tak2022rawboost}.
Furthermore, there have been attempts to find a general representation by fine-tuning speech foundation models \cite{wang2021investigating, eom2022anti} pre-trained on large-scale dataset of speech domain, such as LibriSpeech \cite{panayotov2015librispeech}.
Additionally, attention mechanisms have been introduced to learn discriminative features for anti-spoofing by focusing on spoof-related features and suppressing unrelated ones \cite{ling2021attention, tak2021end}.

Meanwhile, the fundamental distinction between fake and bonafide speech lies in their origin.
Bonafide speech originates from the vocal cords, while fake speech can be generated by a variety of different speech synthesis systems.
In this regard, formulating the ADD task as a binary classification between bonafide and fake speech is impractical because the binary classification method intrinsically assumes that fake speech shares a similar distribution \cite{zhang2021one}.
Since fake speech utterances have different distributions depending on the synthesis system \cite{yan2022initial}, assuming consistent characteristics across all systems is unreasonable.
In addition, rapidly evolving speech synthesis systems are making the distribution of fake speech more diverse.
\begin{figure}[!t]
  \centering
  \begin{minipage}{0.45\linewidth}
    \centering
    \includegraphics[width=\linewidth]{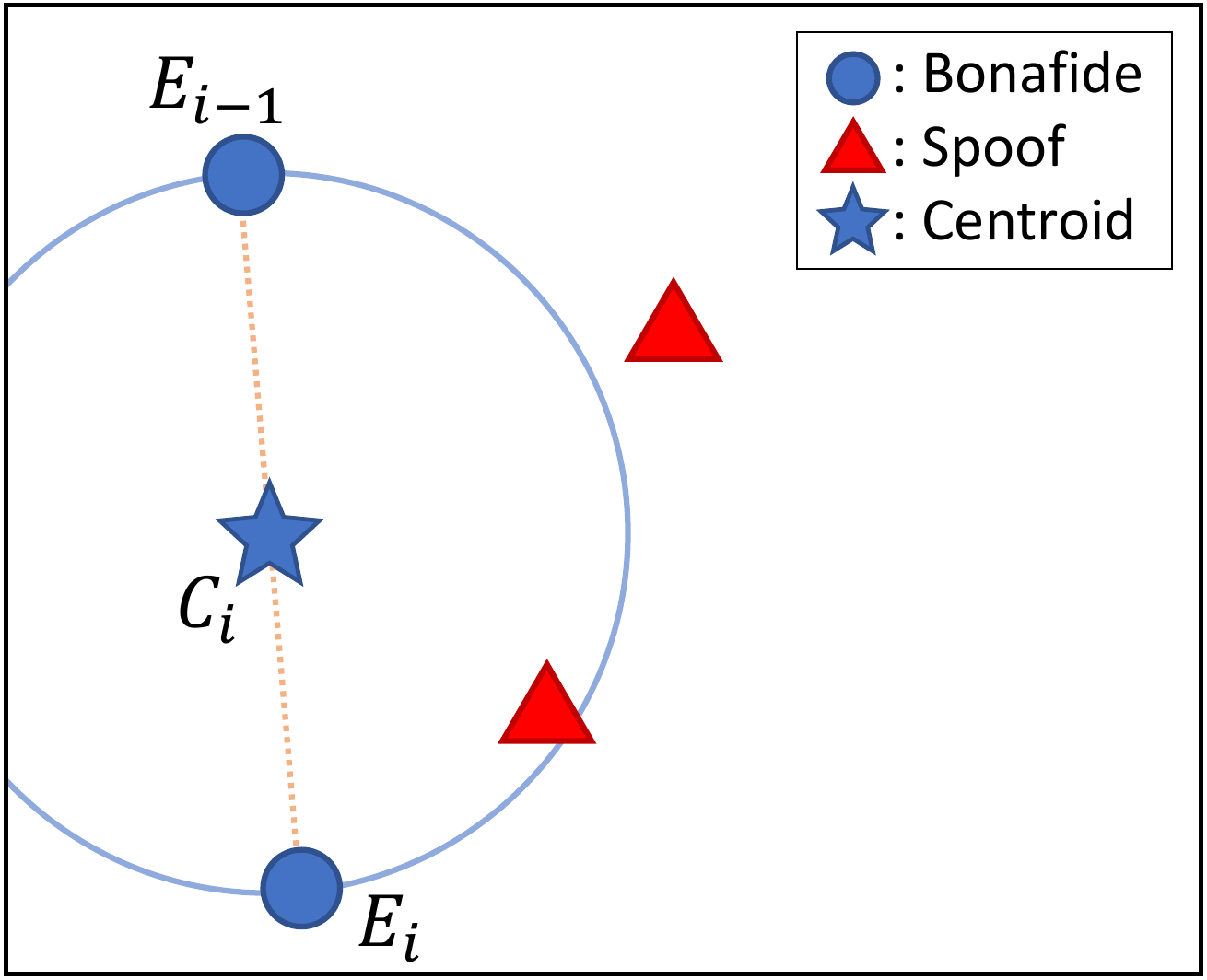}
        $i$-th minibatch
  \end{minipage}%
  \begin{minipage}{0.1\linewidth}
    \centering
    \includegraphics[width=\linewidth]{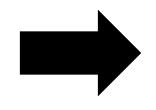}
  \end{minipage}%
  \begin{minipage}{0.45\linewidth}
    \centering
    \includegraphics[width=\linewidth]{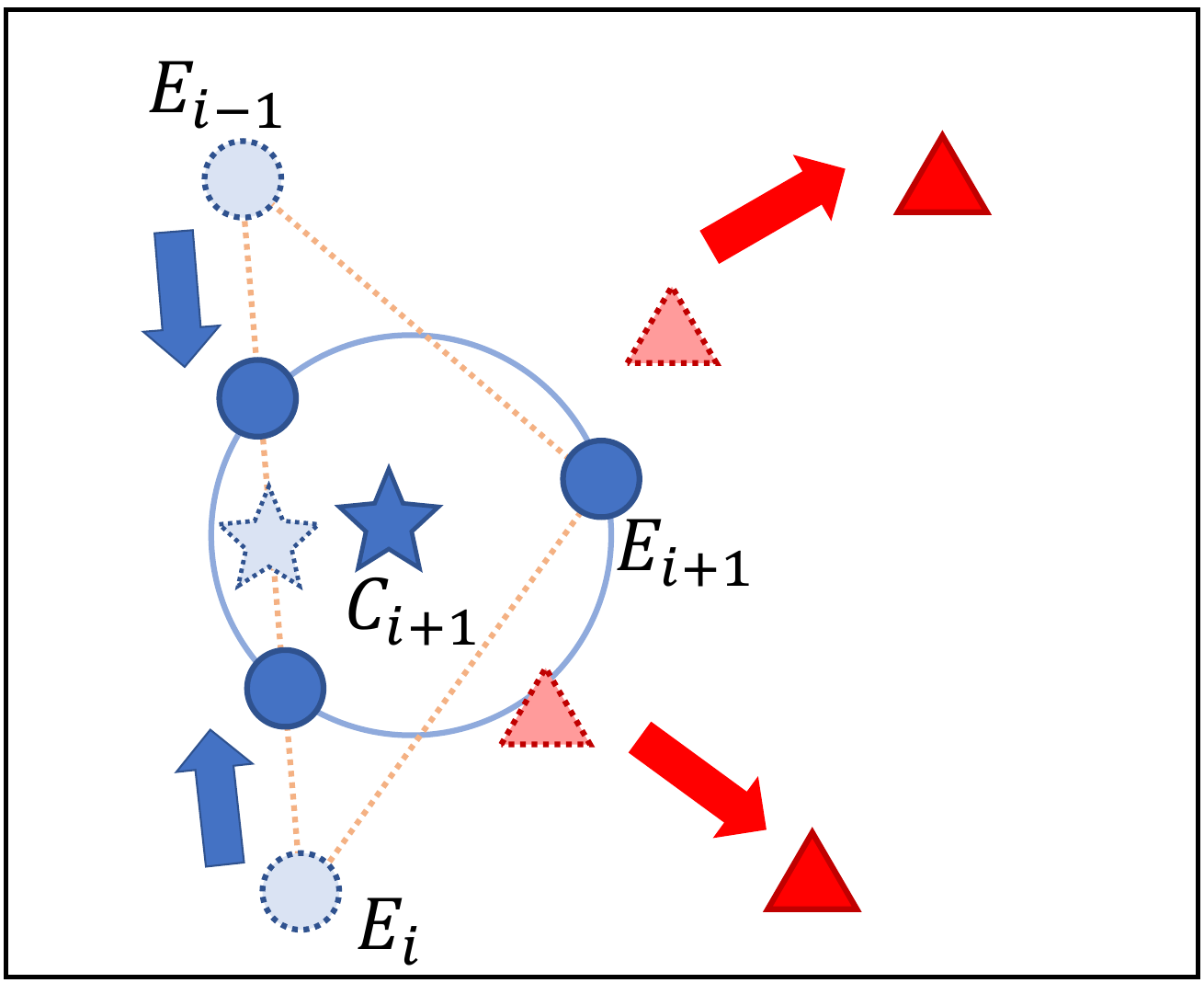}
    $(i+1)$-th minibatch
  \end{minipage}%
  \caption{Illustration of the ACS method when the $(i+1)$-th minibatch is the input and each minibatch contains one bonafide sample. All dashes show their previous states, and the optimization movements are represented by arrows.}
  \label{fig:shift}
\end{figure}

Instead of binary classification, recent studies \cite{zhang2021one, alegre2013one} proposed a one-class classification method \cite{moya1993one} for ADD.
The main idea of the one-class classification method is to learn the distribution of bonafide samples and set an appropriate boundary around them, considering all samples outside the boundary as fake speech.
OC-Softmax \cite{zhang2021one}, a representative one-class method for ADD, sets a tight margin for the bonafide samples, creating a compact boundary around the centroid vector.
However, the centroid is influenced by both classes during the training process.
Considering the important role of the centroid, which represents the bonafide class, its representativeness can be affected by fake samples.

To address this issue, we propose an ACS method, which determines the centroid using only bonafide speech.
During training, the centroid vector is updated as a weighted average of bonafide samples.
Utilizing the centroid obtained through ACS, we apply one-class learning to optimize bonafide samples to move closer to the updated center, while fake samples move further away.
Our method maps the bonafide samples into a single cluster and forms a well-separated and simplified feature space, demonstrating superior generalization ability.
Additionally, we employ the speech foundation model pre-trained on large-scale dataset \cite{wang2021voxpopuli, pratap2020mls} consisting solely of bonafide speech.
This can facilitate the extraction of features of bonafide speech that are highly discriminative from fake speech.

In this work, our proposed method demonstrates its superior generalization ability by outperforming the state-of-the-art (SOTA) system on the ASVspoof 2021 deepfake (DF) and 2019 logical access (LA) datasets.
Furthermore, the t-SNE \cite{van2008visualizing} visualization illustrates that our method effectively maps the bonafide representations into a single cluster and forms a well-separated and simplified feature space.
\section{Method}
\subsection{XLS-R feature encoder}
We employ a speech foundation model, XLS-R \cite{babu2021xls}, as the front-end feature encoder.
XLS-R is a model for self-supervised cross-lingual speech representation learning based on wav2vec 2.0 \cite{baevski2020wav2vec}.
XLS-R is pre-trained on 128 languages and approximately 436k hours of unlabeled speech data, showing strong performance across various downstream speech tasks such as speech translation and speech recognition \cite{babu2021xls}.
Since XLS-R is pre-trained on large-scale bonafide speech data from various languages, XLS-R can capture the inherent features of bonafide speech.
Moreover, XLS-R shows superior anti-spoofing performance compared to other speech foundation models \cite{wang2021investigating}.
Therefore, we utilize XLS-R as the most appropriate front-end feature encoder for ADD.
\subsection{Attentive statistics pooling}
In order to incorporate both local and global level spoofing evidence present in the utterance, we introduce Attentive Statistics Pooling (ASP) \cite{okabe2018attentive}.
Some studies \cite{zhang2023impact, zhang21da_interspeech} have observed a significant performance degradation when the specific time frames are removed.
This implies that evidence of voice spoofing, which helps to distinguish between fake and bonafide utterances, can vary across time frames.
On the other hand, the evidence of voice spoofing may exist not only at the local level but also at the global level, such as excessive smoothing \cite{liu2023leveraging}.

ASP uses an attention mechanism to obtain utterance-level features by assigning weights to each frame.
The utterance embedding is obtained by concatenating the weighted mean and standard deviation vectors.
The weighted mean vector focuses on important frames relevant to spoofing detection, and the standard deviation vector contains spoofing characteristics in terms of temporal variability over long contexts.

Let \({h}_t \in \mathbb{R}^{C}\) denote the frame-level feature at time step $t$.
The attention module computes scalar score $e_t$ for each frame-level feature
\begin{equation}\label{eq:attention}
    \begin{aligned}
        e_t &= v^\top f(W h_t + b)
    \end{aligned}
\end{equation}
where \({v,b} \in \mathbb{R}^{C}\) and \({W} \in \mathbb{R}^{C\times C}\) are learnable parameters and $f(\cdot)$ is a non-linear activation function.
The score is normalized across all time frames to obtain $\alpha_t$.
\begin{equation}
    \begin{aligned}
        \alpha_t &= \frac{\exp(e_t)}{\sum_{\tau} \exp(e_\tau)}
    \end{aligned}
\vspace{-2pt}
\end{equation}
The normalized score $\alpha_t$ represents the importance of each frame relevant to spoofing detection and is reflected as the weight of each frame to compute the weighted mean vector \({\widetilde{\mu}} \in \mathbb{R}^{C}\).
\begin{equation}
    \widetilde{\mu} = \sum_{\tau} \alpha_\tau h_\tau
\vspace{-2pt}
\end{equation}
 The weighted standard deviation \({\widetilde{\sigma}} \in \mathbb{R}^{C}\) is defined as follows:
 \begin{equation}
    \widetilde{\sigma} = \sqrt{\sum_{\tau} \alpha_\tau h_\tau \odot h_\tau - \widetilde{\mu} \odot \widetilde{\mu}}
\vspace{-2pt}
\end{equation}
where $\odot$ represents the element-wise product.
\begin{figure}[!t]
  \centering
  \includegraphics[width=6.9cm]{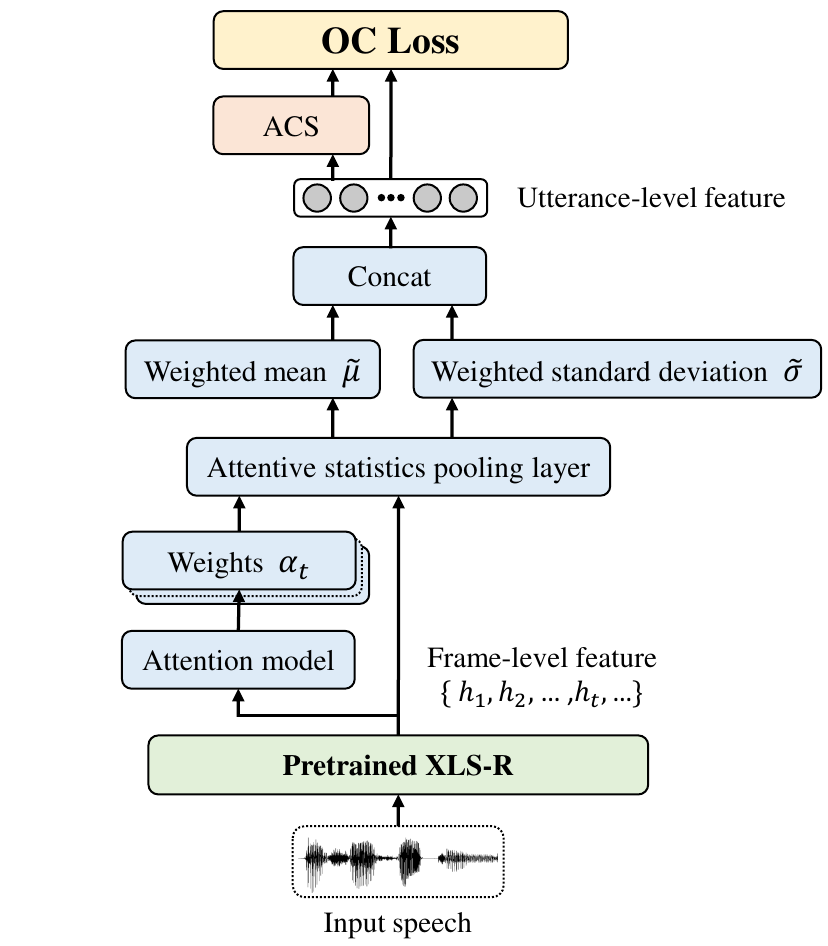}
  \caption{The pipeline of our proposed model. The blue boxes indicate the ASP module. 
  The frame-level feature is extracted from the XLS-R feature encoder, and the utterance-level feature is obtained through the ASP module.}
  \label{fig:main}
\end{figure}

\subsection{One-class learning with Adaptive Centroid Shift}
We use a one-class learning approach rather than binary classification methods.
Binary classification methods intrinsically assume that fake speech has a similar distribution \cite{zhang2021one}, but this assumption is not suitable as fake speech exhibits different distributions \cite{yan2022initial} and continues to evolve.
In one-class learning, the fundamental idea is to map embeddings of the target class closer to each other while pushing embeddings of non-target classes further away.
OC-Softmax \cite{zhang2021one} is a representative one-class learning method for anti-spoofing.
It uses a trainable centroid vector that is influenced by both bonafide and fake samples.
However, the fake samples can affect the representativeness of the centroid characterizing the bonafide class.

In order to obtain the specialized centroid vector representing the bonafide class, we determine the centroid directly with ACS.
The ACS method continuously updates the centroid vector by a weighted average of bonafide samples only when the bonafide samples are present within the mini-batch.
After applying ACS, we optimize the bonafide samples to be closer to the centroid vector while pushing the fake samples away from the centroid vector.
The key point of the ACS method is to define the centroid vector using only bonafide samples.

To elaborate, we initialize the centroid as the first encountered bonafide speech representation.
Then, we continuously update the centroid by calculating the weighted average of the bonafide speech representations.
The ratio of bonafide to fake samples in a single batch is 1 to 9. Including too many bonafide samples in a batch can alter the centroid, leading to decreased training stability.
In Eq (\ref{eq:centroid}), $C_{i}$ denotes the bonafide centroid vector determined by a total of \textit{n} bonafide samples up to the ${i}$-th mini-batch.
When there are \textit{s} bonafide samples in the $({i+1})$-th mini-batch, the ${(i+1)}$-th centroid vector becomes
\begin{equation}\label{eq:centroid}
    {C}_{i+1} = \frac{n{C}_{i} + s{E}_{i+1}}{n+s},
\end{equation}
where \({E}_i \in \mathbb{R}^D \) is the average of bonafide embeddings of ${i}$-th mini-batch.

Once the centroid is defined by ACS for a certain mini-batch, our one-class loss function optimizes bonafide class embeddings to be closer to the centroid and fake class embeddings to be far from the centroid.
We design the intuitive and straightforward one-class loss function based on metric learning.
The one-class loss function $\mathcal{L_{OC}}$, integrated with the centroid $C$ obtained by ACS, is designed with the cosine distance metric, given by
\begin{equation}\label{eq:loss_acs}
    \begin{aligned}
        \mathcal{L_{OC}} = 
        -\frac{1}{M_b}\sum_{i=1}^{M_b}\frac{r_{b,i}^\top C}{||r_{b,i}|| \cdot ||C||}
        +\frac{1}{M_s}\sum_{j=1}^{M_s}\frac{r_{s,j}^\top C}{||r_{s,j}|| \cdot ||C||}
    \end{aligned}
\end{equation}
where \( {r}_s, {r}_b, C \in \mathbb{R}^D \) are respectively a vector of spoof, bonafide and centroid and $||\cdot||$ denotes the computation of the 2-Norm.
$M_b$, $M_s$ are the number of bonafide and fake samples in a mini-batch, respectively.

\section{Experimental setup}
\subsection{Datasets and Metrics}
In all experiments, we trained and validated our models using the train and development partitions of the ASVspoof 2019 LA dataset \cite{wang2020asvspoof}.
We evaluated our method on three subsets to investigate the generalization ability: ASVspoof 2019 LA\,(19LA), ASVspoof 2021 LA\,(21LA), and ASVspoof 2021 DF\,(21DF) \cite{yamagishi2021asvspoof}.
The 21DF dataset consists of 600k utterances and includes more than 100 different spoofing attack algorithms involving audio coding and compression artifacts.
The 19LA evaluation set consists of 13 different TTS and VC systems.
The 21LA uses the same algorithms as 19LA for generating spoofed speech data and also reflects encoding and transmission effects.

We used the EER as the primary evaluation metric, and for the 19LA and 21LA, we also utilized the minimum normalized Tandem Detection Cost Function (min t-DCF) \cite{kinnunen2018t} as the additional metric. 
\subsection{Implementation Details}
\textbf{Data pre-processing}\quad
All audio data are cropped or concatenated to create segments of around 4 seconds duration \cite{tak2022automatic}.
Rawboost \cite{tak2022rawboost} is utilized for data augmentation.
For the 21LA database, we use a combination of linear and non-linear convolutive noise and impulsive signal-dependent additive noise strategies, while for the others, we use stationary signal-independent additive noise with random coloration \cite{tak2022automatic}.\\
\textbf{XLS-R feature encoder}\quad 
We use a pre-trained XLS-R model comprising 0.3B parameters to extract a feature representation from the raw input waveform.
The XLS-R model is implemented by using the fairseq framework \cite{ott2019fairseq}.\\
\textbf{Training details}\quad 
In the training phase, the XLS-R model is jointly optimized with the back-end classifier without freezing.
The output sequences of the feature encoder are passed through the ASP module to obtain the utterance-level embedding.
The Adam optimizer is used with an initial learning rate of $10^{-6}$, a weight decay of $10^{-4}$, and  a batch size of 20. 
With the maximum number of training epochs set to 100, we implement early stopping to prevent over-fitting when the EER on the validation set showed no improvement for 7 consecutive iterations.
The final system is derived by averaging the weights \cite{izmailov2018averaging} of model checkpoints from the top 5 epochs with the highest EER performance on the validation set.
All experiments are conducted on a single GeForce RTX 4090 GPU.
\begin{figure*}[t]
    \centering
    \begin{subfigure}{3.5cm}
          \centering
          \includegraphics[width=\linewidth]{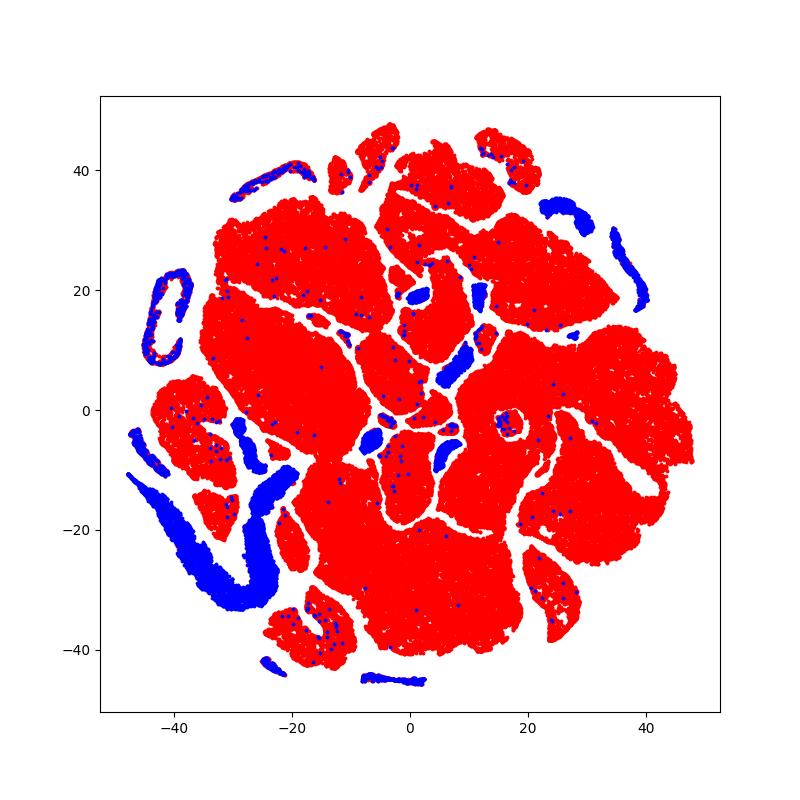}
    \end{subfigure}%
    \hspace{-1.25cm}
    \begin{subfigure}{3.52cm}
      \centering
      \includegraphics[width=0.5\linewidth]{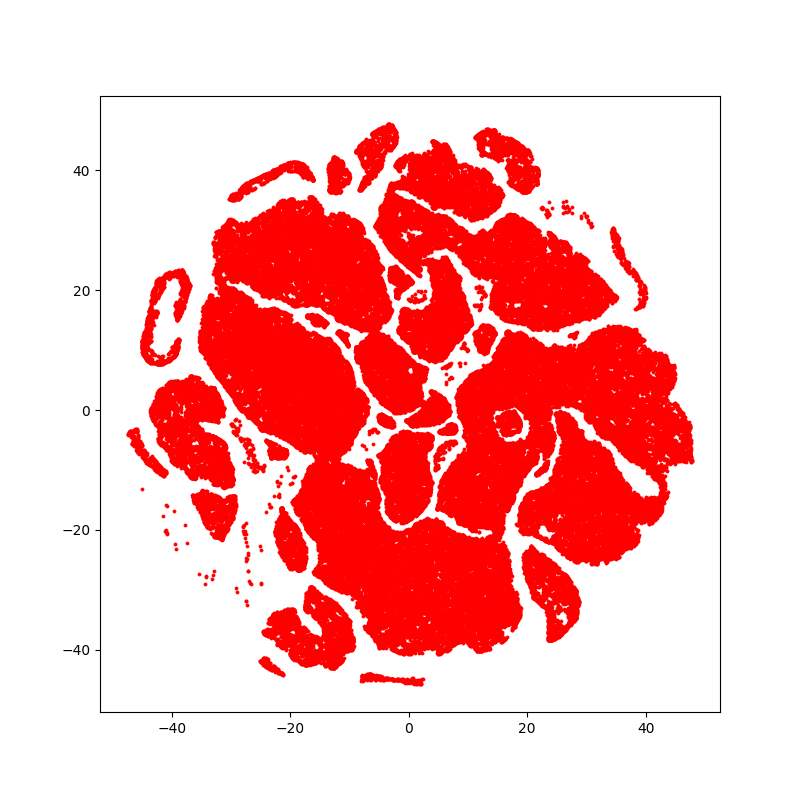}
      \includegraphics[width=0.5\linewidth]{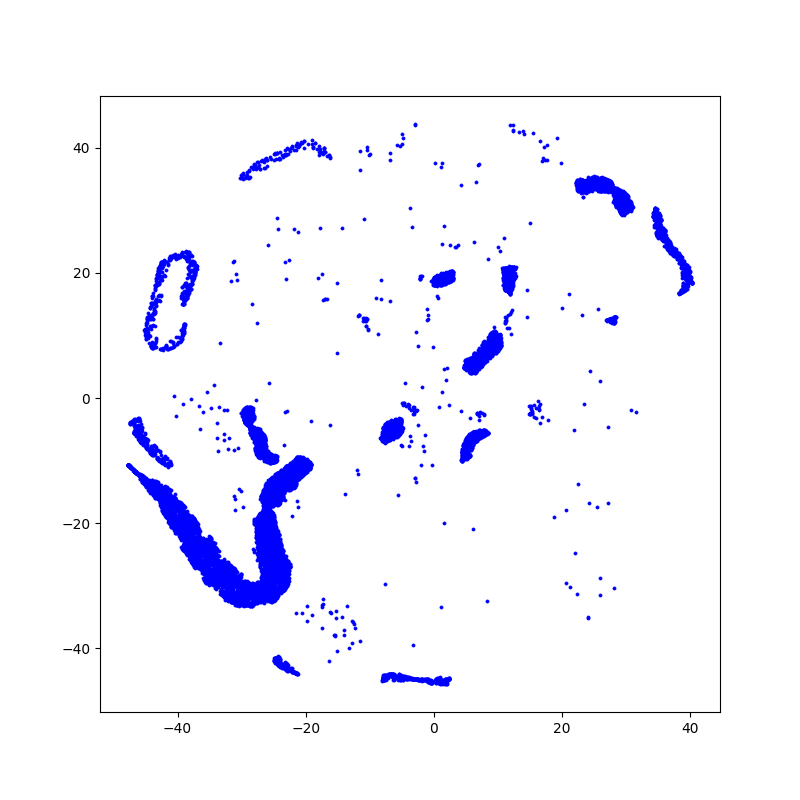}
    \end{subfigure}%
    \hspace{-1.12cm}
    \begin{subfigure}{3.5cm}
      \centering
      \includegraphics[width=\linewidth]{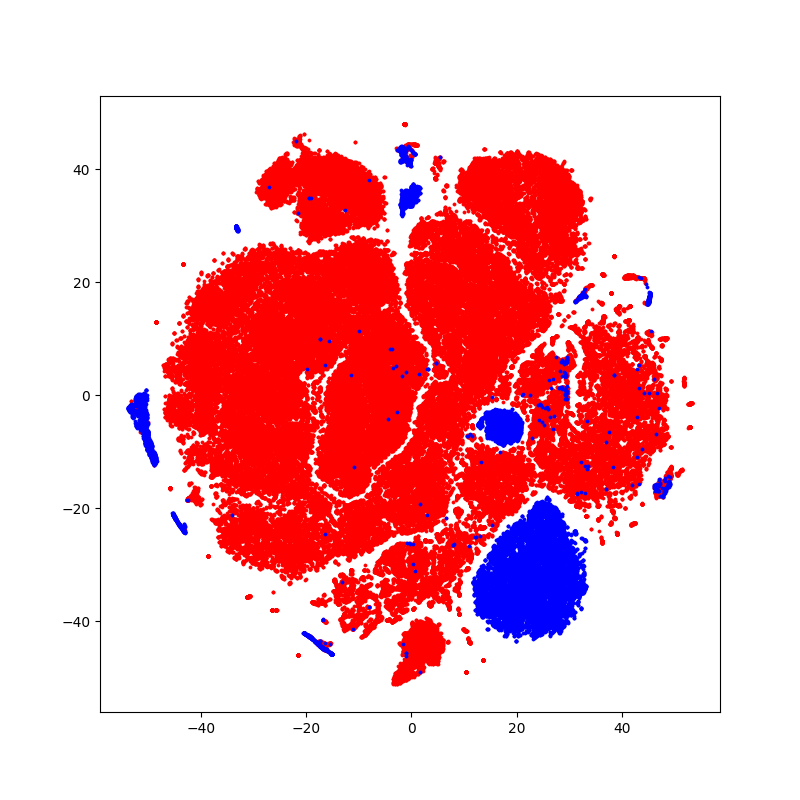}
    \end{subfigure}%
    \hspace{-1.25cm}
    \begin{subfigure}{3.52cm}
      \centering
      \includegraphics[width=0.5\linewidth]{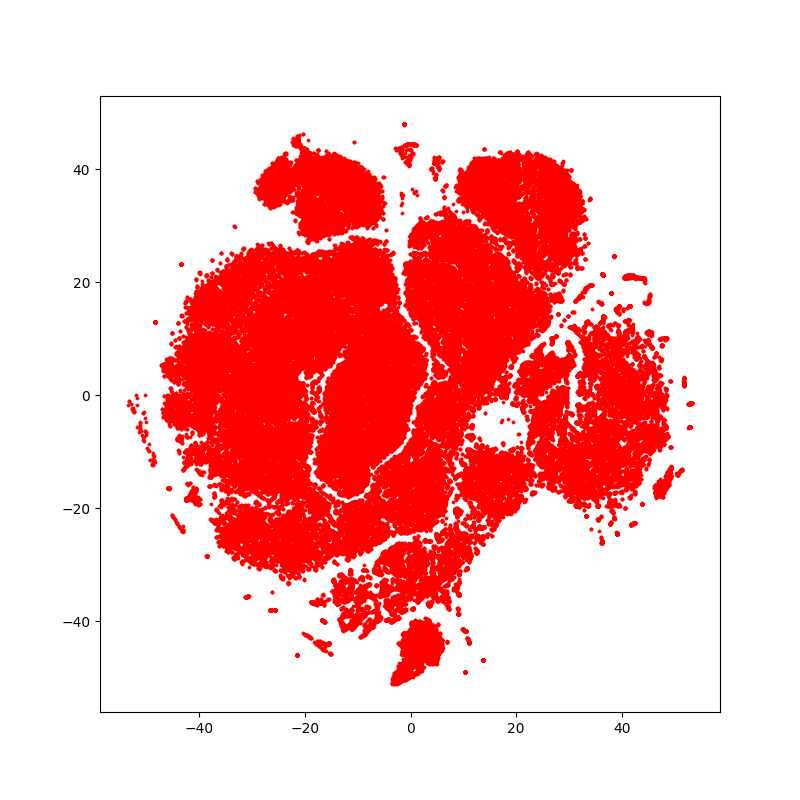}
      \includegraphics[width=0.5\linewidth]{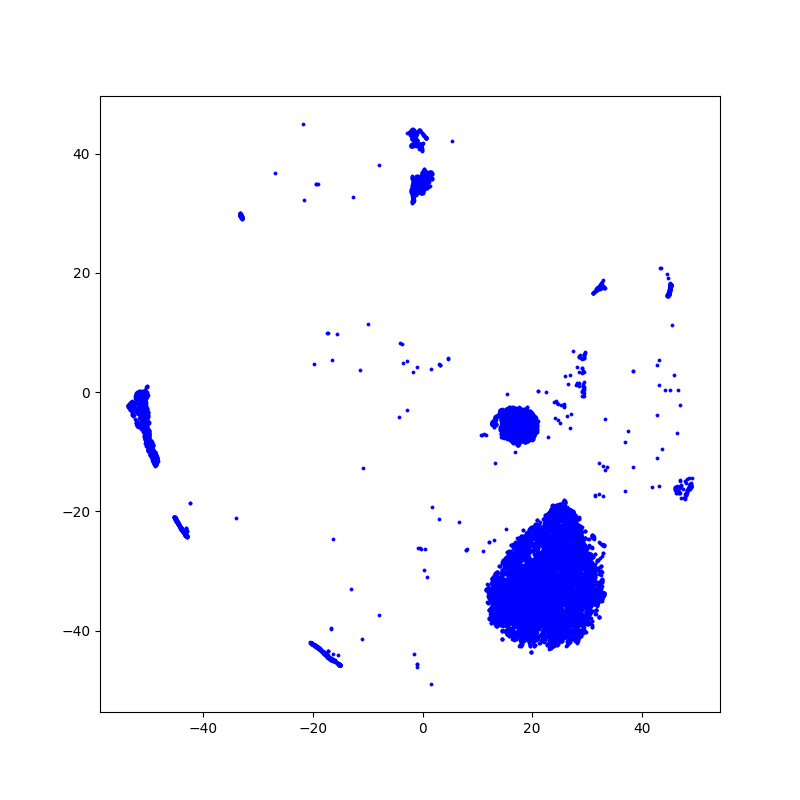}
    \end{subfigure}%
    \hspace{-1.12cm}
    \begin{subfigure}{3.5cm}
      \centering
      \includegraphics[width=\linewidth]{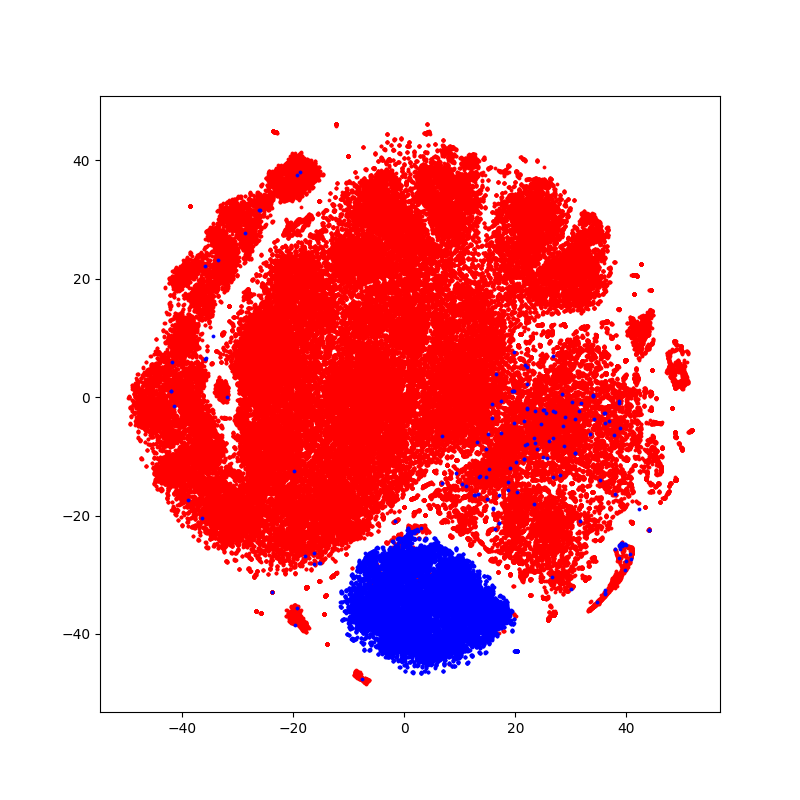}
    \end{subfigure}%
    \hspace{-1.25cm}
    \begin{subfigure}{3.52cm}
      \centering
      \includegraphics[width=0.5\linewidth]{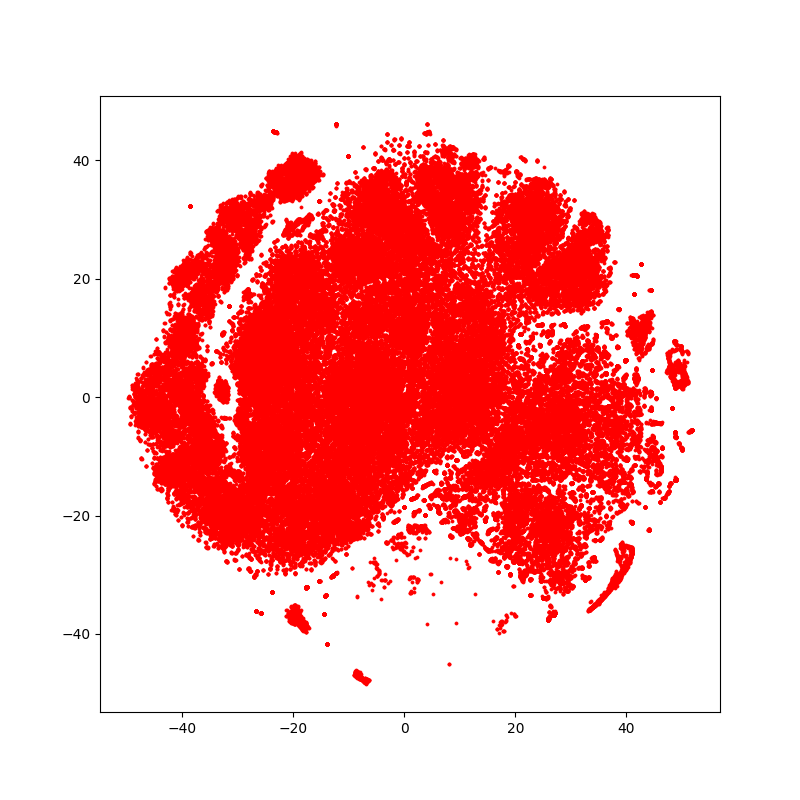}
      \includegraphics[width=0.5\linewidth]{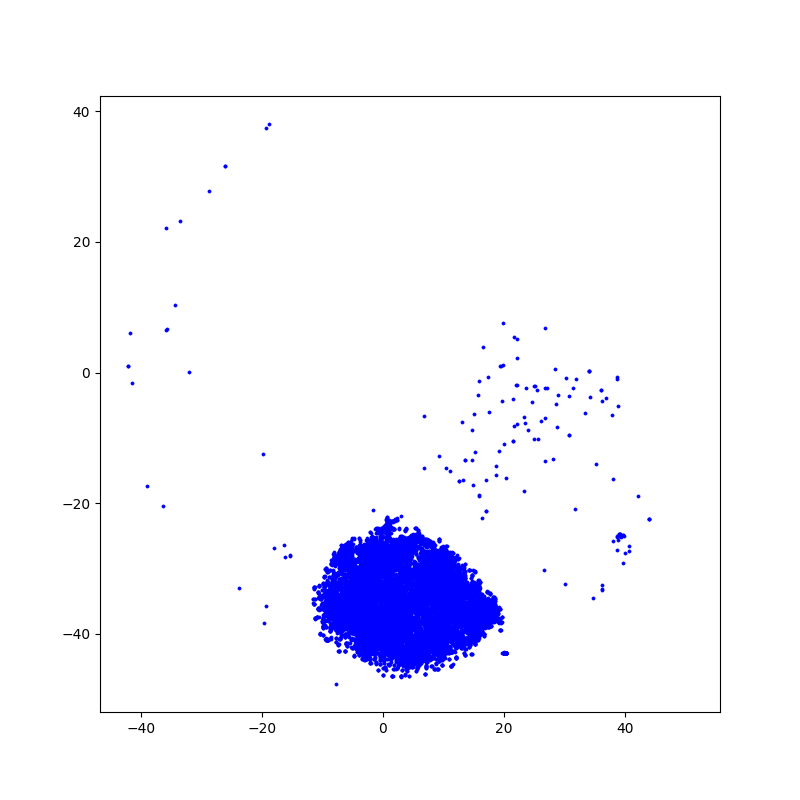}
    \end{subfigure}%
    \hspace{-1cm}
    \begin{subfigure}{2cm}
      \includegraphics[width=\linewidth]{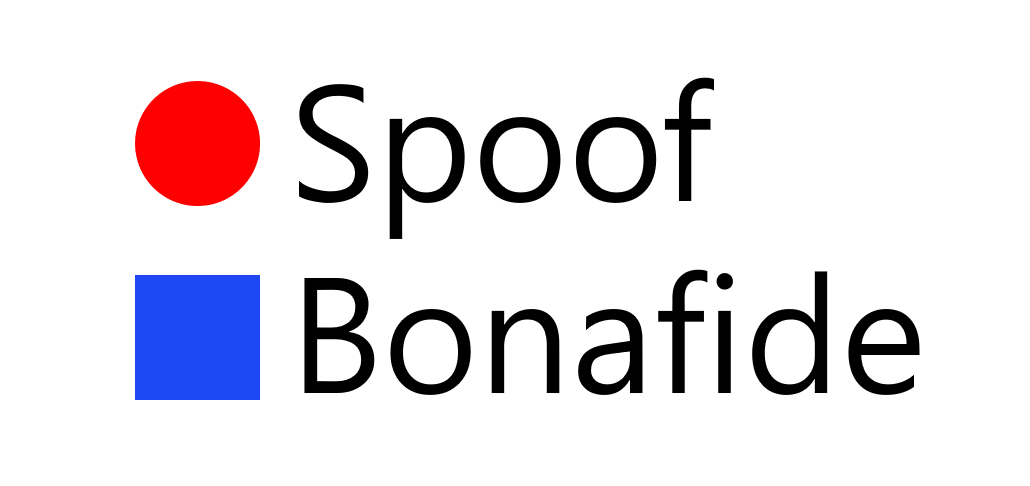}
      \vspace{+1cm}
    \end{subfigure}%

\hspace{-1cm} (a) WCE \hspace{+3cm} (b) OC-softmax \hspace{+3cm} (c) ACS+OC
\caption{Visualization of the embedding space using t-SNE on the ASVspoof 2021 LA evaluation dataset.}
\label{fig:t-SNE embedding}
\end{figure*}
\section{Results and Analysis}
\begin{table}[!t]
  \centering
  \caption{Comparison of our system with other recent systems evaluated on the ASVspoof 2021 DF dataset in terms of EER (\%).}
  \label{tab:2021DF}
  \begin{tabular}{ll}
    \toprule
    \textbf{System}    & \textbf{EER (\%)$\downarrow$} \\
    \midrule
      Hubert+LCNN \cite{wang2021investigating}  & 12.39 \\
      XLS-R+LCNN \cite{wang2021investigating}   & 4.75 \\
      XLS-R+AASIST \cite{tak2022automatic}      & 2.85 \\
      WavLM+MFA \cite{guo2023audio}             & 2.56 \\
      OCKD \cite{lu2023one}                     & 2.27 \\
      \textbf{Ours}     & \textbf{2.19} \\
    \bottomrule
  \end{tabular}
\end{table}

\begin{table}[!t]
  \centering
  \caption{Comparison of our system with other recent systems evaluated on the ASVspoof 2019 LA and 2021 LA datasets, reported in terms of min t-DCF and EER (\%).}
  \label{tab:2019LA}
  \resizebox{\columnwidth}{!}{%
  \begin{tabular}{llll}
    \toprule
    \textbf{Eval Set} &\textbf{System} &\textbf{EER(\%) $\downarrow$} &\textbf{min t-DCF} $\downarrow$\\
    \midrule    
    \multirow{9}{*}{19LA} 
    & SENet \cite{zhang21da_interspeech}       & 1.14  & 0.0368 \\\relax
    & RawGAT-ST \cite{tak2021end}              & 1.06  & 0.0335 \\\relax
    & AASIST  \cite{jung2022aasist}            & 0.83  & 0.0275 \\\relax
    & DFSincNet \cite{huang2023discriminative} & 0.52  & 0.0176 \\
    & WavLM+MFA \cite{guo2023audio}            & 0.42  & -      \\
    & Wav2Vec2+VIB \cite{eom2022anti}          & 0.40  & 0.0107 \\
    & OCKD         \cite{lu2023one}            & 0.39  & -      \\
    & XLS-R+AASIST \cite{lu2023one}            & 0.22  & -      \\
    & \textbf{Ours}                  & \textbf{0.17 }             &\textbf{0.0050 }\\
    \midrule
    \multirow{8}{*}{21LA}
    & RawNet2 \cite{tak2022rawboost}           & 5.31                   &0.3099\\
    & WavLM+MFA \cite{guo2023audio}            & 5.08 & -\\
    & Wav2Vec2+VIB \cite{eom2022anti}          & 4.92                   &-\\
    & DFSincNet \cite{huang2023discriminative} & 3.05                   &0.2601\\\relax
    & LCNN-LSTM \cite{tomilov2021stc}          & 2.21                   &-\\
    & \textbf{Ours}                  & \textbf{1.30 }   &\textbf{0.2172}\\
    & OCKD         \cite{lu2023one}            & 0.90  & -\\
    & XLS-R+AASIST \cite{tak2022automatic}     & 0.82             &0.2066 \\
    \bottomrule
  \end{tabular}%
  }
  \vspace{-10pt}
\end{table}
\subsection{Results}
In Table \ref{tab:2021DF}, we compare our proposed system with other existing systems evaluated on the 21DF dataset.
As shown in Table \ref{tab:2021DF}, our system outperforms all existing systems with an EER of 2.19\%.
Also, all systems except ours utilize a binary classification methods.
This highlights the importance of the one-class learning approach in the ADD task.
Note that OCKD \cite{lu2023one} utilizes one-class learning but employs a teacher model based on binary classification.
Additionally, our system achieves superior performance by using only a simple ASP module, compared to conventional systems that use complicated classifiers.

In Table \ref{tab:2019LA}, we report the min t-DCF and EER of the proposed system on the 19LA and 21LA evaluation subsets, comparing its performance with other existing anti-spoofing systems.
At 19LA, our system achieves an EER of 0.17\% and a min t-DCF of 0.0050, outperforming all existing systems.
At 21LA, our system achieves an EER of 1.30\% and a min t-DCF of 0.2172, demonstrating the competitive performance compared to the recent SOTA system.
In summary, our proposed system shows excellent anti-spoofing performance not only on the 21DF dataset but also on various other datasets. This illustrates its generalization ability against unseen spoofing attacks.
\subsection{Comparison of one-class and binary classification methods} \label{sec:ablation_oneclass}
In order to demonstrate the power of our methodology, we conduct a comparative analysis among our approach, binary classification method and other one-class method.
In ADD task, OC-softmax \cite{zhang2021one} and weighted cross entropy (WCE) are representative loss functions in binary and one-class classification, respectively.
In Table \ref{tab:ablation_oneclass}, we report the anti-spoofing performance of binary and one-class classification methods, including our proposed method.
All training details are the same except for the loss function and a linear layer added to the WCE.
In the one-class method, the utterance embedding is directly used to the loss function, whereas in the WCE, the utterance embedding passes through a linear layer to ensure the final output has a dimension of 2.

As is shown in Table \ref{tab:ablation_oneclass}, the one-class methods outperform the binary classification method.
We can expect that performance degradation may occur due to the binary classification methods assuming that fake speech shares a similar distribution.
Among the one-class methods, ACS+OC shows the best performance, due to its method of determining the centroid.
The contents related to the centroid are described in Sec \ref{sec:ablation_centroid}.
Also, in Fig.\ref{fig:t-SNE embedding}, we visualize embeddings evaluated on the 21LA evaluation partition using t-SNE \cite{van2008visualizing} to compare ours and other methods.
In contrast to WCE and OC-softmax, our proposed method maps bonafide embeddings into a single cluster, thereby demonstrating a clear separation between bonafide and spoof classes.
\subsection{Analysis of centroid definition methods.}\label{sec:ablation_centroid}
We investigate the anti-spoofing performance based on various centroid definition methods to further support our method.
Table \ref{tab:ablation_acs} compares the anti-spoofing performance for four different centroid definition methods.
In the Table \ref{tab:ablation_acs}, we indicate the classes used to define the centroids for each method.

The fixed type defines the centroid as a random value and does not update it.
The partially fixed ACS type defines the centroid in the same way as ACS for the first 5 epochs and then remains unchanged and fixed.
In the trainable type, the neural network learns the centroid by backpropagation from cosine similarity.

According to the Table \ref{tab:ablation_acs}, the fixed type shows the lowest performance. We can infer that the fixed centroid may not adequately represent the bonafide class.
In the partially fixed centroid, the EER decreased only up to the 5-th epoch on the development set and increased after fixing the centroid.
Comparing Trainable with ACS, the ACS approach shows better performance than Trainable.
In ACS, the centroid is influenced only by bonafide samples during the training process and not by fake samples, which can enhance the representativeness of the centroid.
Furthermore, although outliers may be present in bonafide samples, the ACS method can improve the stability of the training by minimizing the impact of outliers on the centroid.
\begin{table}[!t]
  \centering
  \caption{Performance comparison between binary and one-class methods in terms of EER (\%). All systems are evaluated on the ASVspoof 2021 LA and DF partition.}
  \label{tab:ablation_oneclass}
  \begin{tabular}{l c c c c}
    \toprule
    &\multicolumn{2}{c}{\textbf{Method}} &\multicolumn{2}{c}{\textbf{EER(\%)$\downarrow$}} \\ \cmidrule(lr){2-3} \cmidrule(lr){4-5} \textbf{Loss}&\textbf{Binary} &\textbf{One-class} &\textbf{DF} &\textbf{LA} \\    
    \midrule    
      WCE        & \ding{51}     & \ding{55}    & 3.14 & 1.67 \\
      OC-softmax & \ding{55}     & \ding{51}    & 2.48 & 1.55\\
    \textbf{ACS+OC} & \textbf{\ding{55}} & \textbf{\ding{51}} & \textbf{2.19} & \textbf{1.30}  \\
    \bottomrule
  \end{tabular}
\end{table}
\begin{table}[!t]
  \centering
  \caption{Performance comparison between centroid types in terms of EER (\%). All systems are evaluated on the ASVspoof 2021 DF partition.}
  \label{tab:ablation_acs}
  \begin{tabular}{l c c c}
    \toprule
    \textbf{Type} &\textbf{Spoof} &\textbf{Bonafide} &\textbf{EER(\%)$\downarrow$}   \\
    \midrule
                Fixed                   & \ding{55}     & \ding{55}   &3.13 \\
                Trainable               & \ding{51}     & \ding{51}   &2.52 \\
                Partially fixed ACS     & \ding{55}     & \ding{51}   &2.44 \\
                \textbf{ACS} & \textbf{\ding{55}} & \textbf{\ding{51}} & \textbf{2.19}\\
    \bottomrule
  \end{tabular}
\end{table}
\section{Conclusion}

We propose an ACS for one-class learning to enhance the robustness of the model against unseen spoofing attacks.
The ACS method calculates the centroid using only bonafide samples, enhancing the representativeness of the centroid.
By incorporating ACS with our one-class loss function, we optimize the bonafide samples to be close to the centroid and the fake samples to be away from the centroid.
Our approach maps the bonafide speech representations into a single cluster within the embedding space, improving generalization ability against unseen spoofing attacks.
Furthermore, the ASP module effectively captures spoofing artifacts that exist both locally and globally within each utterance unit.
Our system exhibits outstanding generalization ability against unseen spoofing attacks by outperforming all existing systems on the 21DF and 19LA datasets.
To the best of our knowledge, it is the lowest reported EERs for both 21DF and 19LA databases.
\section{Acknowledgements}
This work was supported by Institute of Information \& communications Technology Planning \& Evaluation (IITP) grant funded by the Korea government (MSIT, 2022-0-00653).

\bibliographystyle{IEEEtran}
\bibliography{mybib}

\end{document}